\newcommand{\Cee}{\ensuremath{\mathcal{C}}}

\documentclass[reprint,amsmath,amssymb,aps]{revtex4-2}

\usepackage{tikz}

\begin{document}
\title{Partition
 Function Zeros of Paths and Normalization Zeros of~ASEPS}
\author{Zdzislaw Burda}\email{zdzislaw.burda@agh.edu.pl} 
\affiliation{AGH University of Krakow, Faculty of Physics and Applied Computer Science, \\
al. Mickiewicza 30, 30-059 Krak\'ow, Poland}
\author{Desmond A. Johnston}\email{D.A.Johnston@hw.ac.uk}
\affiliation{School of Mathematical and Computer Sciences, Heriot-Watt University,\\ Riccarton, Edinburgh EH14 4AS, UK }

\begin{abstract}
We exploit the equivalence between the partition function of an adsorbing Dyck walk model and the Asymmetric Simple Exclusion Process (ASEP) normalization to obtain the thermodynamic limit of the locus of the ASEP normalization zeros from a conformal map.  
We discuss the equivalence between this approach and using an electrostatic analogy to determine the locus, both in the case of the ASEP and the random allocation model.    
\end{abstract}

\maketitle

\section{Introduction}

The question of how the non-analyticity in the thermodynamic potential that signals a phase transition could arise from  the corresponding finite-size partition functions in the thermodynamic limit was first elegantly answered by Lee and Yang~\cite{LY,LY2} who considered expanding these in a complex fugacity. 
The appearance of non-analyticity 
was seen to be due to the accumulation of the zeros of the partition functions on the positive real axis of the complex fugacity plane at the transition point.

An appealing way to understand these results is to employ an electrostatic analogy~\cite{BDL, Brazil}, where the zeros are regarded as point charges in two dimensions. A finite size partition function $Z_S(u)$ for a system of size $S$ can be expanded as a polynomial in the complexified fugacity of interest, $u$, 
\begin{equation}
Z_S(u) = \sum_{n=0}^S a_n u^n \; ,
\end{equation}
where the coefficients $a_n$ will all be positive, which can be written in terms of the polynomial zeros $u_i$ as
\begin{equation}
Z_S(u) = A \prod_{n=1}^S (1-u/u_i) \; ,
\end{equation}
where $A$ is some constant.
The corresponding free energy/thermodynamic potential in the thermodynamic limit
 \begin{equation} \label{psiZ}
\psi(u) =  \lim_{S \to \infty} \frac{\ln  Z_S (u)}{S}
\end{equation}
can be written in terms of $\rho(z)$, the local density of zeros
\begin{equation}
\psi(u) =  \int dz \rho(z) \ln ( 1 - u/z)
\end{equation}
and will in general be complex for complex $z$.
If we consider  $\Phi(u) = \Re \psi(u)$
\begin{equation}
  \Phi(u) = \int dz \rho (z) \ln |(1 - u /z)|
\end{equation}
we find
\begin{equation}
\nabla^2 \Phi(u) = \frac{1}{2 \pi} \rho(u)
\end{equation}
so $\Phi(u)$ is the equivalent to the two-dimensional electrostatic potential associated to a distribution of charges $\rho(u)$.

 In the thermodynamic limit, Lee--Yang zeros typically cluster in wedge-shaped regions that pinch to a critical point located on the real axis or coalesce to form a critical line $\gamma$ that separates different phases in the complex fugacity plane. The latter case occurs in the models discussed here. The functional form of $\psi(u)$ may be different on different sides of the critical line, say $\psi_-(u)$ and $\psi_+(u)$, which correspond to different phases, but their real parts $\Re \psi_-(u) = \Re \psi_+(u)$ must be equal to each other for any $u$ on the phase boundary $\gamma$. If the critical curve is parameterized by a real parameter $s$, the line density $\mu(s)$ of the zeros (charge density) on $\gamma$ is obtained from the difference in the 
 imaginary part of $\psi'(u)$ on both sides of the boundary $u = \gamma(s)$ 
\begin{equation} \label{density}
    \mu (s) = \frac{1}{2 \pi i} 
    \Im \left[ \psi'_+(\gamma(s)) - \psi'_-(\gamma(s))\right] \; .
\end{equation}
{More} 
 precisely, the derivatives are calculated as the limiting values 
at the points $\gamma(s) \pm n(s) \delta$ for $\delta \rightarrow 0^+$
approaching the critical curve from both sides. $n(s)$ is a unit vector perpendicular to the curve $\gamma$ and $\delta>0$ is the distance
from this curve.   

At a physical phase transition, which appears at a real fugacity value,  the complex zeros ''pinch'' the transition point from the upper and lower complex half-plane as the system size increases. This mechanism applies widely to statistical mechanical
\mbox{models~\cite{LY, LY2, Fi64, Abe, Suzuki, wjk, bbckk1, bbckk2}} for both field- and temperature-driven transitions (though at least one example is known where the free energy remains analytic in the presence of such accumulating complex zeros~\cite{shlos}). The nature of the transition and critical exponents can be extracted from the finite-size scaling of the zeros.
Typically, these ideas have been applied numerically in finite-sized systems, with partition function zeros extracted from series expansions, transfer matrix calculations, or numerical simulations ~\cite{lottsashrock}. Analytically determining the thermodynamic limit of the locus of zeros and their density has only been possible in a limited number of models (unsurprisingly, since it is equivalent to exactly solving the models in question)~\cite{1d1, 1d2, 1d3, MF1, IPZ,BGP,prz, REM,stau, fatfish,LY, LY2, Fi64}.

In \cite{us} the locus of zeros and the line density of zeros on the phase boundary
were calculated in the thermodynamic limit for a class of random allocation models exhibiting a condensation transition by what appears, at first sight, to be a rather different approach drawing from analytic combinatorics.
The locus of zeros was found to be simply given by the image of the unit circle under a conformal transformation determined by the generating function of the statistical weights in the model.  

In the current paper, we first summarize the general approach espoused in \cite{FS} and elsewhere \cite{God1, God2,  God-Luck, corb1} to calculating the thermodynamic limit of the free energy/partition function in various models using such analytic combinatorial methods. We briefly review the application of these methods to the canonical and grand canonical random allocation models. The form of the generating function for the grand canonical model allows the locus of partition function zeros for these to be read off directly.

We then apply the methods of \cite{us} to calculate the locus of zeros in a class of lattice path models and, noting the equivalence of their partition functions to the normalization of the ASEP (Asymmetric Simple Exclusion Process), obtain the locus of the ASEP normalization zeros, which had been obtained numerically in \cite{bePRL} and using the ``physical''/electrostatic  approach in \cite{bwjk}. We note that the results on
Lee--Yang zeros of the ASEP normalization obtained within different frameworks are in perfect agreement.

\section{General Observations \label{sect:FS}}

The weighty Flajolet and Sedgewick tome ``Analytic Combinatorics'' \cite{FS} is a veritable treasure trove of analytical methods applied to combinatorial enumeration problems. To quote the book's manifesto, as stated in its preface, ``Analytic Combinatorics aims at predicting the properties of large
structured combinatorial configurations precisely through an approach based extensively on
analytic methods. Generating functions are the central objects of study of the theory''. A similar perspective was taken in the comprehensive studies by Godr\`eche \cite{God1,God2} of condensation and extremes for fixed and fluctuating numbers of random variables and in works such as \cite{God-Luck, corb1}.

To set the scene, we first recall some of the relevant results from \cite{FS} with regard to the generating functions and analytical methods. A central role  is played by the (ordinary) generating function for some set $\{ f_n \}$ of combinatorial objects of size $n$
\begin{equation}
f(z) = \sum_{n} f_n z^n \; .
\end{equation}
{The} 
 coefficient $f_n$ in the formal power series, denoted by $[z^n] f(z)$, can be extracted by using a contour integral if $f(z)$ is viewed as a complex function which is analytic near the origin
\begin{equation}
[z^n] f(z) = f_n = \oint_{|z|=\epsilon} \frac{dz}{2\pi i} \frac{f(z)}{z^{n+1}} 
\label{fn}
\end{equation}
{The}
 evaluation of the integral for a large $n$ can be facilitated in many cases by exponentiating the integrand and then using  saddle point methods 
\begin{equation}
[z^n] f(z) = f_n = \oint_{|z|=\epsilon} \frac{dz}{2\pi i} \exp [ \ln f(z) - (n+1) z ] \; .
\label{fnsaddle}
\end{equation}
{Observation} 
 of various combinatorial structures suggests that in many cases the asymptotic behavior of the coefficients $f_n$ is given by
\begin{equation}
    [z^n] f(z) \sim A^n \theta(n)
\end{equation}
i.e., exponential growth $A^n$ and a subexponential factor $\theta(n)$. It is further observed that
\begin{itemize}
    \item{} The location of $f(z)$'s singularities determine $A$. Specifically, if the closest singularity to the origin along the real axis is $\sigma$, then $A=1/\sigma$
    \item{} The nature of the singularities determine $\theta(n)$.
\end{itemize}
{Additionally,} 
 it is often useful to rescale so that the generating functions are singular at $1$~using
\begin{equation}
    [z^n] f(z) = \sigma^{-n} [z^n] f(\sigma z)  \; .
\end{equation}
{Under} 
 suitable conditions \cite{FS}, it is possible to replace  an $f(z)$ singular at  $1$ with a local approximation near the singularity $s(z) \sim (1-z)^{a_1} \log^{a_2} (1-z)$ such that
\begin{equation}
    [z^n] f(z) \sim [z^n] s(z)
\end{equation}
which allows the asymptotic behavior of the $f_n$ to be extracted from estimating $f(z)$ around the singular point(s) and performing the contour integral in Equation \eqref{fn}.

In a similar manner to $f(z)$, where $z^n$ marks objects of size $n$ in the power series $f(z) = \sum_n f_n z^n$ it is possible to define bivariate generating functions
\begin{equation}
f(u,z) = \sum_{n,k} \,  f_{n,k} \, z^n u^k
\end{equation}
where $u^k$ now additionally marks some scalar parameter with value $k$. A particular case of this is given by the functional composition 
$F(u,z) = g(u f(z))$.  
Assume  that $f,g$ are analytic at the origin  and let $z_F$, $z_f$ and $z_g$ be the radii of convergence of $F, f$ and $g$. If we define the values of $f$ and $g$ at these points to be $\tau_f = f(z_f)$, $\tau_g = g(z_g)$ then (again with suitable conditions on $f$ and $g$) there are three possible regimes to consider:
\begin{itemize}
\item{}{\it Supercritical} $u\tau_f>z_g$. As $z$ is increased from $0$ there will be some value $\tilde z$ strictly less than $z_f$ such that $u f({\tilde z}) =z_g$. In this case, the singularity type is that of the external function $g$.
\item{}{\it Subcritical} $u\tau_f<z_g$. In this case, the singularity of the composition is driven by that of the internal function $f$. 
\item{}{\it Critical} $u \tau_f = z_g$. Here, there is a confluence of the two singularities.
\end{itemize}
In the sequel, we apply these methods first to the random allocation model and then to lattice path models/ASEPs.

\section{Random Allocation Model: Canonical Ensemble}
The random allocation model in the canonical ensemble is  a particular example of a partition function given by a product form with a constraint where analytic combinatorial methods may be gainfully employed.
The model consists of weighted distributions of $S$ indistinguishable particles over $N$ boxes, whose partition function is simply given by~\cite{Ritort, bbj, Ehrenfest, bb, fr}
\begin{equation}
    Z_{S,N} = {\sum}_{(s_1,\ldots,s_N)} 
    w(s_1) \ldots w(s_N) \delta_{S- (s_1+\ldots +s_N)} ,
    \label{ZSN}
\end{equation}
where the Kronecker delta constraint enforces the correct total number of particles.
The single box weights $w(s)$ are non-negative and for convenience when considering power law weights,  $w(s)= 1/s^{\beta}$, we shall take  $s=1,2,\ldots$ We will consider the thermodynamic limit $S,N \to \infty$  with the ratio $S/N = \rho$ fixed, where $\rho$ is the mean density of particles. 

Consider the generating function for the weights in Equation \eqref{ZSN}
\begin{equation}
f(z) = {\sum}_{s=1}^\infty w(s) z^s \; .
\label{f}
\end{equation}
{Following} 
 Equation \eqref{fn}, we can extract $w(s)$ from $f(z)$ by carrying out the appropriate contour integral
\begin{equation}
w(s) = \oint_{|z|=\epsilon} \frac{dz}{2\pi i} \frac{f(z)}{z^{s+1}} 
\label{wint}
\end{equation}
where a suitably small radius $\epsilon>0$ for the circular contour about the origin will ensure that the function $f(z)$ is holomorphic on $|z|<\epsilon$ and only the $w(s)z^s$ term is picked out by the contour integration.

This approach may be immediately extended to evaluate the canonical partition function itself using the contour integral
\begin{equation}
Z_{S,N} = \oint_{|z|=\epsilon} \frac{dz}{2\pi i} \frac{f(z)^N}{z^{S+1}} \; .
\label{oint}
\end{equation}
{A} 
 term in the expansion of $f(z)^N$ of the form $w(s_1)z^{s_1} \cdot w(s_2)z^{s_2} \cdots w(s_N)z^{s_N}$ will only contribute to the integral if $s_1+s_2+\cdots s_N = S$, thus  enforcing the constraint. The expansion of $f(z)^N = ({\sum}_{s=1}^\infty w(s) z^s)^N$  ensures that all the products $\prod_{i=1}^N w(s_i)$ with permissible $s_i$ assignments are included in the expression for the partition function.
For instance, for the power-law weights 
$w(s) = s^{-\beta}$ with  $\beta \in (1,\infty)$ the generating function (\ref{f}) is the polylogarithm $f(z) = {\rm Li}_\beta(z)$.

We can now use  saddle point methods to extract the asymptotic behavior in the thermodynamic limit $N \rightarrow \infty$ with $S/N = \rho$ in Equation \eqref{oint}. If we exponentiate the terms in the integrand in Equation \eqref{oint} and neglect those terms of $O(1/N)$, taking $(S+1)/N \sim S/N$, we find
\begin{equation}
Z_{S,N} = \oint \frac{dz}{2\pi i} \exp[ N (\ln f(z) - \rho \ln z)]
\label{oint2}
\end{equation}
which may be evaluated by saddle point methods to give
\begin{equation}
Z \sim \exp [ N \phi (\rho)]
\label{phi}
\end{equation}
where
\begin{equation}
\phi (\rho) =  -\ln z^*(\rho)\rho+ \ln f( z^*(\rho))
\label{SP1}
\end{equation}
and
\begin{equation}
z^* \frac{f'(z^*)}{f(z^*)} = \rho \; .
\label{SP2}
\end{equation}
{The} 
 phase transition manifests itself as a breakdown in the saddle point solution. As $\rho$ is increased the saddle point value of $z$, i.e., $z^*(\rho)$, also increases and for some families of weights reaches the radius of convergence, typically scaled to be $1$, of the series for $f(z)$ at a finite critical density $\rho_c$. This is, indeed, the case for the power law distributions $w(s) = 1/s^{\beta}$ where the critical
density is $\rho_{\rm c} = \zeta(\beta-1)/\zeta(\beta)$ for $\beta>2$, where $\zeta$ is the Riemann zeta function. 

It is possible to use an exact recursion relation for $Z_{S,N}$
\begin{equation}
    Z_{S,N}=\sum_{s=1}^{S-N+1}w(s) Z_{S-s,N-1}
\label{eq:Z-recurrence}    
\end{equation}
for $S\ge N\geq 1$, and
\begin{equation}
Z_{S,0} =\delta_S 
\label{ZS0}
\end{equation}
for $N=0$ to show that the breakdown of the saddle point equation signals a condensation transition. When $\rho>\rho_c$ the excess balls condense into a single box.

\section{Random Allocation Model: Other Ensembles
\label{sec:GC}}

So far we have discussed the $(S,N)$ ensemble, where the number of
particles $S$ and the number of boxes $N$ are fixed, but
we can consider three other ensembles: $(z,N)$, 
where $N$ is fixed but $S$ is variable; $(S,u)$ 
where $S$ is fixed but $N$ is variable; $(z,u)$ 
where both $S$ and $N$ are variable. They have the corresponding
partition functions
\begin{equation} \label{Ztilde}
    \tilde{Z}_N(z) = \sum_{S=N}^\infty Z_{S,N} z^S,
\end{equation}
\begin{equation} 
    Z_{S}(u) = \sum_{N=1}^S Z_{S,N} u^N.
    \label{ZSmu}
\end{equation}
and 
\begin{equation} \label{F}
    F(z,u)  = 1+\sum_{N=1}^\infty \sum_{S=1}^\infty Z_{S,N} z^S u^N 
\end{equation}
respectively. For convenience, we added the term $1$ before the sum.
As a constant, it does not play any physical role and
can be interpreted as a vacuum contribution $Z_{S,0}=\delta_{S}$ \eqref{ZS0}. 
The fugacity $z$ can be interpreted as an exponent $z=e^{-\mu}$
of the chemical potential $\mu$, and if $N$
is interpreted as the volume of the system, $u$ can be interpreted
as an exponent $u=e^{-p}$ of pressure $p$. 
Using the explicit form of the coefficients $Z_{S,N}$ \eqref{ZSN} in Equation \eqref{Ztilde}) 
we see~that 
\begin{equation} 
    \tilde{Z}_N(z) = f(z)^N,
\end{equation}
and hence 
\begin{equation}\label{Ff}
    F(z,u) = 1+\sum_{N=1}^\infty f(z)^N u^N = \frac{1}{1- u f(z)} .
\end{equation}
{On} 
 the other hand, it follows from (\ref{F}) that
\begin{equation}
    F(z,u) = 1+\sum_{S=1}^\infty Z_S(u) z^S 
\end{equation}
so 
\begin{eqnarray}
    Z_S(u) &=& \oint_{|z|=\epsilon}\frac{dz}{2\pi iz^{S+1}} F(z,u) \nonumber \\
 &=&
    \oint_{|z|=\epsilon}\frac{dz}{2\pi iz^{S+1}}\frac{1}{1-u f(z)}
    \label{uzeros}
\end{eqnarray}
where $\epsilon$ must be small enough not to capture any additional
singularity outside the origin of the complex plane.

We now focus on the partition function $Z_S(u)$ of the $(S,u)$-ensemble~\cite{God2, bbj2, bbje, bbjr, bbjz}, viewed as a polynomial
in $u$ \eqref{ZSmu}. In particular, we are interested in the locus of
zeros, $Z_S(u_j)=0$, $j=1,\ldots,S$ in the complex plane in the limit $S\rightarrow \infty$. Paradoxically, it is easier to determine the locus of the zeros of the polynomial $Z_S(u)$ not by solving the problem
directly in the $u$ variable, but by examining the asymptotic properties of the integrand \eqref{uzeros} in the integral variable $z$. As follows from the general discussion on electrostatic analogy in the Introduction, in the limit $S\rightarrow \infty$, when the zeros coalesce, they form a critical line at which the derivative of the thermodynamic potential $\psi'(u)$ has a discontinuity. 
The magnitude of the discontinuity is related to the density of zeros on this line \eqref{density}. The critical line can be derived from the asymptotic properties of the integral \eqref{uzeros} as
a borderline (branch cut of $\psi(u)$) between two different solutions $\psi_1(u)$ and $\psi_2(u)$ \cite{us}. If the function $z\rightarrow f(z)$ is injective within the radius of convergence $\sigma$, that is for $|z|<\sigma$, the critical line and thus also the limiting locus as $S \to \infty$ of the Lee--Yang zeros in the complex $u$ plane is given by a remarkably simple formula \cite{us}
\begin{equation}
u = \gamma(s) = \frac{1}{f(\sigma e^{i s} )} \, , \; \; 0 < s \le 2 \pi \; .
\label{fmap}
\end{equation}
{In} 
 other words, the critical line is the image of the boundary of the convergence disc (circle or radius $\sigma$) under the conformal mapping $z \rightarrow u = 1/f(z)$. The critical line $\gamma$ arises at the interface of two different asymptotic regimes of the integral \eqref{uzeros}. One regime corresponds to the situation where the position of the pole $1/(1-uf(z))$ \eqref{uzeros} viewed from the perspective of the variable $z$
\begin{equation}
z_0(u)=f^{-1}(1/u) 
\label{pole}
\end{equation}
is in the convergence disc, that is for $u$ outside the critical curve $\gamma$. In this case, the main contribution to the integral comes from the saddle point $z_{SP}(u)$, 
which lies very close to $z_0(u)$ inside the convergence disc. 
The distance $|z_{SP}(u)-z_0(u)|$ tends to zero as $S\rightarrow \infty$,
so in the limit one can replace $z_{SP}(u)$ by $z_0(u)$. The leading contribution to the integral $Z_S(u)$ is
$\exp(S (z_0(u) + o(1)))$. 
The second regime is for $u$ inside the critical line. In this case the position $z_0(u)$ of the pole $1/(1-u f(z))$
is outside the convergence disc. 
The two cases give the result
\begin{equation}
   \label{psi_solution}
    \psi(u)=\left\{\begin{array}{ll} \psi_1(u) = - {\rm ln} \left(f^{-1}(1/u)\right), & {\rm for} \ u \ {\rm outside} \ \gamma,\\
    \psi_2(u) = -\ln \sigma , & {\rm for} \ u \ {\rm inside} \ \gamma. \end{array}\right.
\end{equation}
{The} 
 saddle point solution $\psi_1(u)$ corresponds to the fluid phase, while $\psi_2(u)$ corresponds to the
condensed phase. This result is consistent with the result that was first derived for real fugacity $u>0$ \cite{bbje}. 
The real parts of the solution are continuous on the critical curve, while the imaginary parts
are not. The discontinuity $\Im \psi'_2(u) - \Im \psi'_1(u)$ gives 
the density of zeros on the critical line \eqref{density}. The density of zeros
on the critical line $u=\gamma(s)$ can be calculated from the
equation \cite{us}
\begin{equation} \label{mus}
    \mu(u) du = \frac{1}{2\pi} ds
\end{equation}
for $s\in (0,2\pi]$, which just reflects the fact that the density on $\gamma$
is an image of uniform density on the circle under the conformal map \eqref{fmap}.

These results may also be derived in the general framework of \cite{FS} 
discussed in \mbox{Section \ref{sect:FS}} for the functional composition of the generating functions
\begin{equation}
    g(z) = \frac{1}{1-z}
\end{equation}
and $f(z)$ to obtain $F(z,u) = g(u f(z))$. In this language, the condensed phase corresponds to the supercritical regime of the composition (the pole is the dominant singularity) and the fluid phase corresponds to the subcritical regime (the singularity of $f(z)$ is dominant).

\section{Binomial Weights and Lattice Paths \label{sect:paths}}

\begin{figure}[h]
\vspace{0.5cm} 
\begin{tikzpicture}[scale=0.6]

\draw[->] (0,0) -- (12,0) ;

\draw[thick] 
    (0,0) -- (1,1) 
    -- (2,0) -- (3,1) 
    -- (4,2) -- (5,1) 
    -- (6,0) --(8,2) -- (9,1) --(10,2)--(12,0);

\filldraw[fill=black] (0,0) circle (0.07); 
\filldraw[fill=black] (6,0) circle (0.07); 
\filldraw[fill=black] (12,0) circle (0.07); 

\filldraw[fill=black] (2,0) circle (0.07); 

\node[above right] at (-0.2,0.1) {\small $c$};
\node[above right] at (1.8,0.1) {\small $c$};
\node[above right] at (5.8,0.1) {\small $c$};
\node[above right] at (11.8,0.1) {\small $c$};
\node[above right] at (0.8,0.1) {\small $u$};
\node[above right] at (3.8,0.1) {\small $u$};
\node[above right] at (8.8,0.1) {\small $u$};

\end{tikzpicture}
\vspace{0.5cm} 
\caption{A Dyck 
 walk with contact fugacity $c$, and additionally
with excursion fugacity $u$.}\label{fig1}
\end{figure}
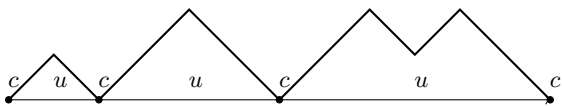

In this section, we recall the problem of enumeration
of Dyck walks, which are lattice paths in the plane that start and end on the horizontal line ($x$-axis), do not go below it
and consist of diagonal steps $(1,1)$ or $(1,-1)$. 
This problem can be mapped onto the random allocation
model. This map can, in particular, be used to apply the machinery developed
for this model to determine Yang--Lee zeros for Dyck walks, see  Figure \ref{fig1}. 
The generating function for Dyck walks with a fugacity $c$ for each contact with the horizontal axis is  given by \cite{jvr}
\begin{equation}
\label{Dyck}
G_D (z,c) =  \frac{c} { 1 - c f_D(z)}= \frac{c} { 1 - c ( 1 - \sqrt{1 - 4 z}) /2 }\;,
\end{equation}
in which $z^{1/2}$ is the fugacity for each step. This can
be obtained by noting that $f_D(z)=( 1 - \sqrt{1 - 4 z}) /2$
is the generating function for a single excursion above the
horizontal axis so $G_D(z,c)$ concatenates multiple excursions with a contact weight of $c$ every time they hit the horizontal axis. If we
use the fugacity $u$ which is associated with the number of excursions between the contacts rather than the fugacity $c$ associated
with the number of contacts, then the generating function \eqref{Dyck} 
will take the form \eqref{Ff} that we discussed for the random
allocation model
\begin{equation}
\label{FDyck}
F_D (z,u) =  \frac{1} { 1 - u f_D(z)}\;,
\end{equation}
because the number of excursions is one less than the
number of contacts. The corresponding weights \eqref{f} for 
$f_D(z)=( 1 - \sqrt{1 - 4 z}) /2$ are 
\begin{equation} \label{wDyck}
    w_D(s) = \frac{1}{2} 4^s (-1)^{s+1} \binom{1/2}{s} = 
    \frac{1}{2(2s-1)} \binom{2s}{s} .
\end{equation}
{They}
 count the number of possible shapes of a single excursion
consisting of $2s$ steps.

The weights for the Dyck walks \eqref{wDyck} 
are a special case of binomial weights 
\begin{equation}
w(s) = (-1)^{s+1} \frac{a}{\sigma^s} \binom{\theta}{s}
\end{equation}
for $\theta \in (0,1)$, $a,\sigma \in (0,\infty)$ 
and $s = 1,2,...$ considered in \cite{us}. 
The corresponding generating function is
\begin{equation}
    f(z) = a\left(1-\left(1-\frac{z}{\sigma}\right)^{\theta}\right)
\end{equation}
{This} 
 function can be explicitly inverted for $|z|<\sigma$. The inverse 
function is
\begin{equation}
    f^{-1}(z) = \sigma\left(1-\left(1-\frac{z}{a}\right)^{1/\theta}\right) .
\end{equation}
{For} 
 Dyck walks  \eqref{wDyck} the parameters are $\theta=1/2$, $a=1/2$ and $\sigma=1/4$:
\begin{equation}
    \label{fDyck}
    f_D(z) = \frac{1}{2} \left(1-\sqrt{1-4z}\right)
\end{equation}
and
\begin{equation}
    f_D^{-1}(z) = \frac{1}{4} \left(1-\left(1- 2z\right)^2\right) .
\end{equation}

\begin{figure}[h]
    \centering
    \includegraphics[width=0.6\linewidth]{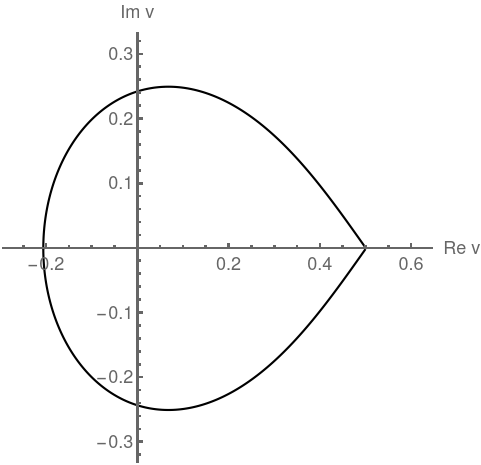}
    \caption{The analytically calculated locus of zeros for adsorbing Dyck walks in the complex $v$ plane. The transition at $v_{cr}=1/2$ is clearly visible, as is its second-order nature, since the impact angle of the locus of zeros is $\pm 3\pi/4$.}
    \label{fig:dyck}
\end{figure}

In the $(S,u)$-ensemble, the Dyck walks have two phases, depending
on the parameter $u$. In the thermodynamic limit $S\rightarrow \infty$,
the phase transition is at $u_{cr} = 1/f_D(1/4) = 2$.
For $u>u_{cr}$ the sum \eqref{ZSmu} is dominated by large values of
$N$, that is, many short excursions, while for $u<u_{cr}$ by small $N$, that is, fewer but longer excursions. Actually, one of them is much longer than the others. The equivalence with the random allocation model allows one to easily understand the results in the language of random allocations \cite{bbje} but also to extend the analysis to the complex plane
to calculate the locus of the zeros of the partition function
$Z_S(u)$ for the Dyck walks for $S\rightarrow \infty$ using the
method described in Section \ref{sec:GC}. We will study the locus of the Yang--Lee 
zeros in the variable $v=1/u$ rather than in $u$, due to the relation to the ASEP
model that will be discussed later. Of course, the descriptions in $v$ and $u$ are equivalent. In the $v$ plane the critical curve $s\rightarrow \gamma(s)$
on which zeros are located is
\eqref{fmap}
\begin{equation}
\label{vgamma}
v = \gamma(s) = f_D(e^{i s}/4) = \frac{1}{2} \left( 1 - \sqrt{1 - e^{i s}} \right)
\end{equation}
for $0 < s \le 2 \pi$. It is plotted in Figure \ref{fig:dyck}. 
Inside this curve, $\psi_D(v)$ is given by \eqref{psi_solution}
\begin{equation} \label{psid}
    \psi_D(v) = -\ln \left( \frac{1}{4} \left(1 - \left(1- 2v \right)^2\right)\right) = -\ln\left[ v(1-v) \right]
\end{equation}
while outside it, $\psi_D(v) = \ln 4 $. 
Note that the interior of the critical curve in the $v$ plane corresponds
to the exterior of the corresponding critical curve in the $u$ plane, and vice~versa. 

The phase transition point at $v_{cr}=1/2$ is clearly visible in Figure \ref{fig:dyck}. 
Assume that $v$ is real. In this case, \eqref{psiZ} and \eqref{ZSmu} entail
\begin{equation}
  -v \psi'_D(v) = \lim_{S\rightarrow \infty} \frac{\langle N \rangle_{S,v}}{S}
\end{equation}
where 
\begin{equation}
   \langle N \rangle_{S,v} = \frac{1}{Z_S(v)} \sum_{N=1}^S N Z_{S,N} v^{-N} 
\end{equation}
is the average number of excursions for a given $S$ and $v$. 
The inverse excursion length in the thermodynamic limit is
\begin{equation}
m(v) = \lim_{S\rightarrow \infty} \frac{\langle N \rangle_{S,v}}{S} = 
\left\{ 
\begin{array}{ll} \frac{1 -2v}{1-v} & \mbox{for} \ v \le 1/2 \\
0 & \mbox{for} \ v > 1/2
\end{array}
\right. .
\end{equation}
{We}
 see that $m(v)$ is the order parameter of the phase transition between the short and long excursion phases. The phase transition is continuous,
as $m(v)$ is continuous at the critical point $v_{cr}=1/2$.
Taking the next derivative we obtain a measure of fluctuations
\begin{eqnarray}
-v m'(v) &=&  \lim_{S\rightarrow \infty} \frac{\langle N^2 \rangle_{S,v} - \langle N \rangle^2_{S,v}}{S} \nonumber \\
&=& 
\left\{ 
\begin{array}{ll} \frac{v}{(-1+v)^2} & \mbox{for} \ v \le 1/2 \\
0 & \mbox{for} \ v > 1/2
\end{array}
\right.
\end{eqnarray}
which is discontinuous at $v_{cr}=1/2$, so the phase transition
is of the second order. 

Returning to the complex plane, let us see how these critical properties are
reflected in the distribution of zeros of the partition function. We are interested
in the vicinity of the point $v_{cr}=1/2$ on the complex plane. 
Denote $\Delta v = v - v_{cr}$ for points on the critical curve $v=\gamma(s)$ for
small $s>0$, $s\ll 1$. Equation \eqref{vgamma} gives 
\begin{equation} \label{vs}
  \Delta v = e^{i 3\pi/4} \sqrt{s} + o(\sqrt{s}) 
\end{equation}
{The} 
 impact angle of the critical line on the real axis at this point is $3 \pi/4$
in the upper half plane and $-3 \pi/4$ in the lower half plane. 
Note that the corresponding angles in the $u$-plane are $\mp \pi /4$ (in the lower
and upper half-planes, respectively). The calculated values $\pm 3 \pi/4$ are
a particular case of a general solution for the random allocation model \cite{us} 
where the impact angles depend on the transition order (and the critical exponents). 
Impact angles in the range $(\pi/2,3 \pi/4]$ in the upper half-plane correspond to second-order phase transitions. The angle $3\pi/4$ is a boundary between the second- and third-order phase transitions.

The distribution of zeros near the critical point can be calculated from Equation \eqref{mus}.
If we parametrize the density on the critical curve $\gamma$ by the distance from the critical point, then Equation \eqref{mus} takes the form
$\mu (|\Delta v|) d|\Delta v|= ds/(2\pi)$ from which we deduce for
$|\Delta v| \ll 1$ that $|\Delta v| \approx \sqrt{s}$ \eqref{vs} and
\begin{equation} \label{muv}
    \mu(|\Delta v|) = \frac{1}{2\pi} \frac{ds}{d|\Delta v|} \approx \frac{|\Delta v|}{\pi}
\end{equation}
so the density of zero goes linearly to zero, which means that zeros are sparser near the critical point.

To conclude this section, as a side remark we note that if the Dyck walks are allowed to move into the lower half plane the corresponding generating function is given by \cite{jvr}
\begin{equation}
\label{Dyck2}
\tilde{F}_{D} (u, z) = \frac{1} { 1 - u \tilde{f}_D(z) }
\end{equation}
where $\tilde{f}_D(z) = 2 f_D(z) = 1 - \sqrt{1-4z}$. The factor
reflects the fact that there are twice as many types of elementary excursions as in the previous case because each type of excursion in
the upper plane has a mirrored equivalent in the lower plane, so
$\tilde{w}_D(s) = 2 w_D(s)$. This type of Dyck path is sometimes referred to as a Bernoulli bridge \cite{God2}. If a probabilistic approach is used directly instead of enumerating  Bernoulli bridges, noting that in each step the path may go up or down with probability $1/2$ gives
the corresponding probabilistic weights \eqref{wDyck} 
\begin{equation}
    \hat{w}_D(s) = \frac{1}{2^{2s}} \tilde{w}_D(s) = 
    \frac{2}{2^{2s}} w_D(s) = 
    (-)^{s+1} \binom{1/2}{s}
\end{equation}
and thus the corresponding generating function is 
$\hat{f}_D(z)= 1 - \sqrt{1-z}$.
\vspace{-6pt}

\section{ASEPs and Lattice Paths}
\label{sec:ASEP}

Blythe and Evans \cite{bePRL,Brazil} were the first to observe that the Yang--Lee zeros approach to  equilibrium phase transitions could also be applied to the  phase transitions in the non-equilibrium steady states of the one-dimensional (totally) Asymmetric Simple Exclusion Process, or ASEP. 
This is defined by particles hopping from left to right on a one-dimensional lattice with open boundaries.
The continuous-time variant of the model is
specified by three rates:\\

\begin{small}
\begin{tabular}{lc}
\textbf{Move}                                 & \textbf{Rate} \\
{Particle} 
 inserted onto empty left boundary site & $\alpha$      \\
Particle removed from occupied right boundary site  & $\beta$       \\
Particle hops to empty site on right         & $1$          
\end{tabular}
\end{small}\\

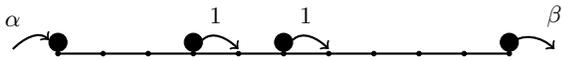
\begin{figure}[h]
\vspace{0.25cm} 
\begin{center}
\begin{tikzpicture}[scale=0.6]

\draw[thick] (0,0) -- (10,0);

\foreach \x in {0,1,...,10} {
    \draw[fill=black] (\x,0) circle (0.05); 
}

\draw[fill=black] (3,0.25) circle (0.2); 
\draw[fill=black] (5,0.25) circle (0.2); 
\draw[fill=black] (0,0.25) circle (0.2); 
\draw[fill=black] (10,0.25) circle (0.2); 

\draw[->,thick] (-1.0,0.1)  node[left,above=5pt] {\small $\alpha$} to[out=45,in=135] (-0.2,0.3); 
\draw[->,thick] (10,0.2) to[out=45,in=135] (11.0,0.1) node[left,above=5pt] {\small $\beta$}; 

\draw[->,thick] (3.2,0.3) node[right=5pt,above=3pt] {\small $1$} to[out=45,in=135] (4.0,0.1) ;
\draw[->,thick] (5.2,0.3) node[right=5pt,above=3pt] {\small $1$} to[out=45,in=135] (6.0,0.1) ;

\end{tikzpicture}
\end{center}
\vspace{0.25cm} 
\caption{(Totally) {Asymmetric} 
 Exclusion Process on a line with insertion rate $\alpha$ and removal rate $\beta$. The internal jump rate to the right is $1$ and only single site occupancy is allowed.}\label{fig3}
\end{figure}

\noindent
For the jump rate $\lambda$, the probability that a jump occurs in an infinitesimal
time interval $\Delta t$ is $\lambda \Delta t$. Additionally, only one particle may occupy a lattice site, see  Figure \ref{fig3}. 

\noindent
The
 quantity of interest in the ASEP is  $Z$ (suggestively named to recall a partition function), which  normalises the statistical weight, $w(\Cee)$, 
of some particle configuration, $\Cee$, in the non-equilibrium steady states. This is given by
\begin{equation} \label{Zgen}
Z = \sum_{\Cee} w(\Cee)\;,
\end{equation}
thus the normalized probability of being in state
$\Cee$ is $P(\Cee)=w(\Cee)/Z$.
The weights $w(\Cee)$ are obtained through the stationarity condition
on the transition rates $W(\Cee \to \Cee^\prime)$ which specify the model
\begin{equation}
\label{steadystate}
\sum_{\Cee^\prime \ne \Cee} \left[ w(\Cee^\prime) W(\Cee^\prime \to
\Cee) - w(\Cee) W(\Cee \to \Cee^\prime) \right] = 0
\;.
\end{equation}
$W(\Cee \to \Cee^\prime)$ is the probability of making the transition from configuration $\Cee$ to $\Cee^\prime$ in a single timestep.
Note that this is less restrictive than the detailed balance
condition for equilibrium states, for which there is a term-by-term
cancellation in the sum in Equation~\eqref{steadystate}.

\begin{figure}[h]
\begin{center}
\begin{tikzpicture}[scale=4, line width=0.8pt]
    \draw[->] (0, 0) -- (1.0, 0) node[anchor=north] {$\alpha$};
    \draw[->] (0, 0) -- (0, 1.0) node[anchor=east] {$\beta$};
    
    \draw[thick,dashed] (0, 0) -- (0.5, 0.5) node[midway, below right, rotate=45] {\small $\alpha = \beta$};

    \draw[thick] (0.5, 0.5) -- (1, 0.5); 
    \draw[thick] (0.5, 0.5) -- (0.5, 1); 

    \node at (0.25, 0.75) {\small (1)};
    \node at (0.75, 0.25) {\small (2)};
    \node at (0.75, 0.75) {\small (3)};
    
    \node[anchor=north east] at (0.6, 0) {\small $\frac{1}{2}$};
    \node[anchor=south west] at (-0.15, 0.45) {\small $\frac{1}{2}$};

\end{tikzpicture}
\end{center}
\caption{The ASEP phase diagram. Region (1) is a low-density phase, region (2) is a high-density phase and region (3) is the maximal current phase. The dotted transition line along $\alpha=\beta<1/2$ is first order and the $\alpha=1/2, \beta=1/2$ transition lines are second order. }
\label{ASEPphase}
\end{figure}
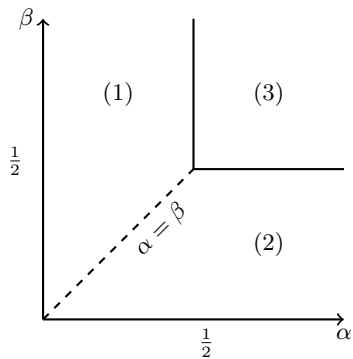

The solution of the ASEP  \cite{DEHP} made use of a matrix product ansatz
to calculate $Z_N(\alpha,\beta)$, the normalization for $N$ sites
\begin{equation}
\label{asepZ}
Z_N (\alpha,\beta) = \sum_{p=1}^{N} \frac{p(2N-1-p)!}{N!(N-p)!}  \frac{
(1/\beta)^{p+1} - (1/\alpha)^{p+1} }{ (1/\beta)-(1/\alpha) } \; .
\end{equation}
{Introducing} 
 the generating function
of the $Z_N (\alpha,\beta)$, or the ``grand-canonical''
normalization, ${\cal Z}(z, \alpha,\beta) = \sum_N Z_N (\alpha,\beta)  z^N$
and performing the summation then gives \cite{bwjk}
\begin{eqnarray}
 {\cal Z} ( z , \alpha, \beta) &=& \frac{1}{ \left(1 - \frac{1}{\alpha} f_D (z)\right)
 \left(1 -  \frac{1}{\beta} f_D(z)\right)} \nonumber \\
&=& 
 F_D\left(z,\frac{1}{\alpha}\right) F_D\left(z,\frac{1}{\beta}\right)
\label{xaxb} 
\end{eqnarray}
where $f_D(z)= (1 - \sqrt{1 - 4 z }) /2$, which clearly
links the ASEP normalization to the partition function of Dyck walks
\eqref{FDyck} and random allocation model. Establishing this equivalence
was crucial for understanding the ASEP phase diagram. The phase diagram
is shown in Figure \ref{ASEPphase}. It comprises low- and high-density phases and a maximal current phase. There is a first-order transition line between the low- and high-density phases for
$\alpha=\beta<1/2$ and second-order transition lines between the low- and high-density and maximal current phases. Blythe and Evans found that this phase structure was reflected in the properties of the zeros of $Z_N (\alpha, \beta)$ in the complex $\alpha$ or $\beta$ planes, exactly as in equilibrium models. Subsequently, it was realized that ASEP normalization (a non-equilibrium quantity) was equivalent to the partition function (an equilibrium quantity) for a class of lattice paths, so the usual Yang--Lee mechanism is in operation \cite{Brazil, bwjk, bwjk2, wbe}.

The product form of this expression shows that the grand-canonical ASEP normalization is identical to the partition function for an
ensemble of pairs of non-interacting Dyck walks, with the excursion fugacities $u_1=1/\alpha$ and $u_2=1/\beta$
(or their inverses $v_1=\alpha$, $v_2=\beta)$. 
We now mirror the discussion on obtaining the Yang--Lee zero locus from \mbox{Equation \eqref{uzeros}} for the adsorbing Dyck walks in
the case of ASEP. Using \eqref{xaxb} we have
\begin{equation}
    Z_N (\alpha,\beta)=\oint_{|z|=\epsilon}\frac{dz}{2\pi iz^{N+1}} \frac{1}{ (1 -  \frac{f_D (z)}{\alpha} ) (1 -  \frac{ f_D(z)}{\beta})} 
    \label{abzeros}
\end{equation}
Polynomials $Z_N (\alpha,\beta)$ are symmetric in $\alpha$ and $\beta$: $Z_N (\alpha,\beta) = Z_N(\beta,\alpha)$. We will
analyze them as polynomials in $\alpha$ for fixed $\beta$.
We are interested in the location of the phase transition
for real $\alpha$, and the distribution of the Yang--Lee 
zeros associated with this transition. The integrand
\eqref{abzeros} has a branch point singularity coming from
$f_D(z)= (1 - \sqrt{1 - 4 z }) /2$ at $|z| = 1/4$, which leads
to a singularity of the thermodynamic potential in the
thermodynamic limit at $\alpha=1/2$ and a pole when $\alpha \rightarrow \beta$, coming from $1/(1-\beta^{-1}f_D(z))$, 
which leads to a singularity at $\alpha=\beta$. The closest singularity to the origin determines the phase so a transition occurs when two are equal, i.e.,  at $\alpha_{cr}(\beta)=1/2$ 
for $\beta\ge 1/2$ or $\alpha_{cr}(\beta)=\beta$ for $\beta<1/2$.

\begin{figure}[h]
  \centering
    \includegraphics[width=0.6\linewidth]{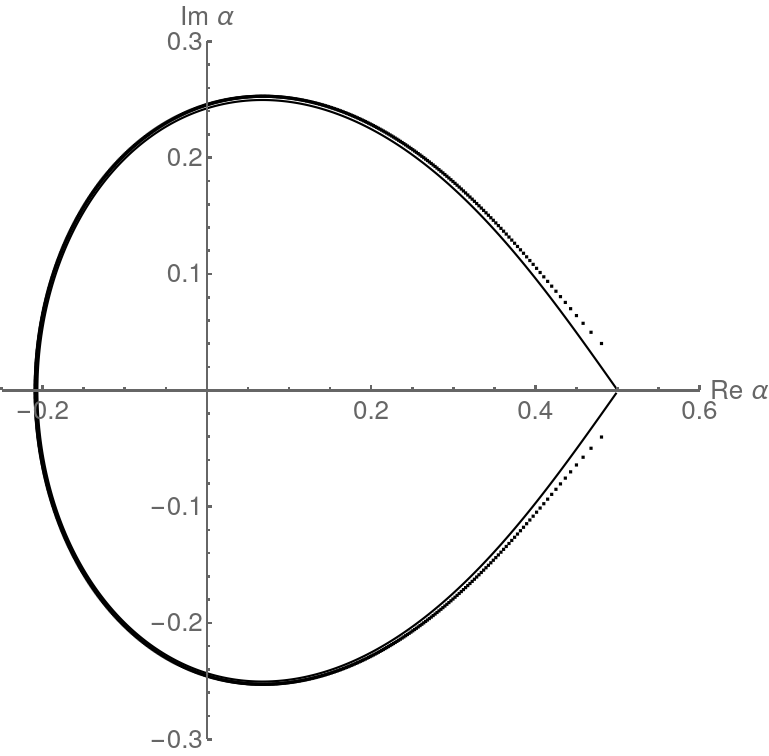}
    \caption{Analytically calculated locus of zeros in the thermodynamic limit from ${\cal Z} (z, \alpha,\beta)$ and numerically determined zeros from $Z_N$ with  $N=1000$ for the ASEP normalization with $\beta=3/4$ (the second-order transition regime). The analytic locus is identical to that for the adsorbing Dyck walks in Figure \ref{fig:dyck}.}
    \label{fig:beta34}
\end{figure}

Taking into account the square root singularity first, the locus of zeros will be given by $\alpha = f_D(\frac{e^{i s}}{4})$,
for $s\in (0,2\pi]$, when $\beta\ge 1/2$. For example, 
looking at $\beta=3/4$ for  the complex zeros in the variable $\alpha$, we can compare the zeros calculated numerically from $Z_N (\alpha,\beta)$ in Equation \eqref{asepZ} with this locus. 
We can see in Figure \ref{fig:beta34} that there is already a reasonable agreement for $N=1000$ and as $N$ is increased, the zeros approach the analytically determined locus further.
The term $1/(1-\beta^{-1}f_D(z))$ does not affect the limiting distribution of zeros in this case.
The phase transition point at $\alpha=1/2$ appears clearly on the real axis, where the two lines of zeros meet at a right angle. Since we effectively repeat the analysis of the adsorbing Dyck walks, the impact angle of the zeros is again $\pm 3\pi/4$, as befits a second-order transition. The density of zeros vanishes at the transition point $\alpha=1/2$ (for $\beta>1/2$), as expected for a second-order transition \eqref{muv}.
The locus of zeros in the complex $\beta$ plane will be identical
if we swap the roles of $\alpha$ and $\beta$.

\begin{figure}[h]
    \centering
    \includegraphics[width=0.6\linewidth]{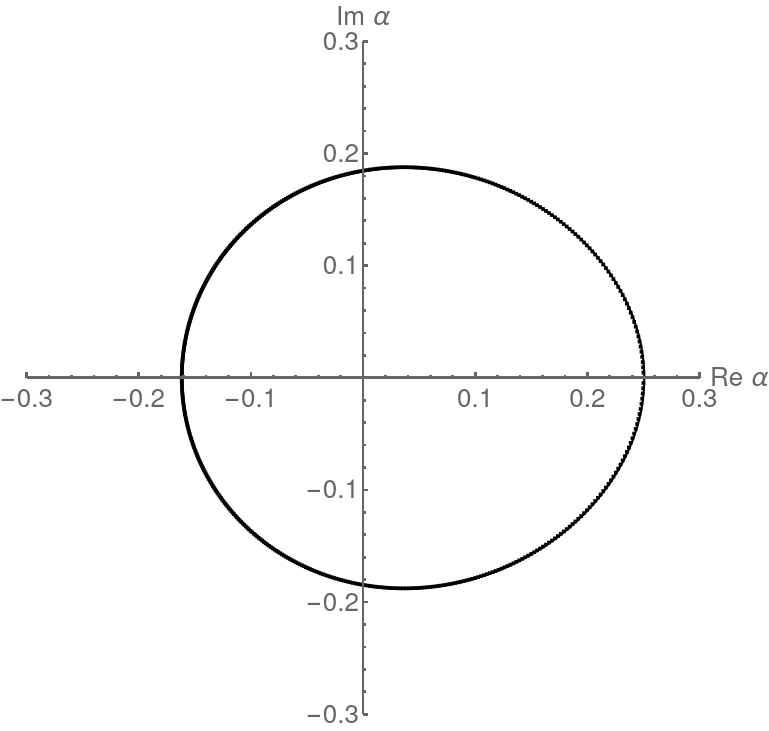}
    \caption{Analytically calculated locus of zeros in the thermodynamic limit from ${\cal Z} (z, \alpha,\beta)$ and numerically determined zeros from $Z_N$ with  $N=1000$ for the ASEP normalization with $\beta=1/4$ (the first-order transition regime). The two are largely indistinguishable.}
    \label{fig:beta14}
\end{figure}

As mentioned, for $\beta<1/2$, the dominant singularity in the variable $\alpha$ is when $\alpha\rightarrow \beta$, since it comes from the pole $1/(1- \beta^{-1} f_D(z))$.
Thus, $z$ is a solution of
$\beta=f_D(z)$, which can be inverted to give $z=\beta(1-\beta)=\alpha(1-\alpha)$.
The critical curve $\gamma$ that gives the locus of zeros
in the thermodynamic limit associated with this singularity is the $f_D$ image of the circle of radius $\beta(1-\beta)$:
$\alpha = \gamma(s) = f_D\left(\beta(1-\beta) e^{is}\right)$, for $s\in (0,2\pi]$.
As an example,
in Figure \ref{fig:beta14} we show the results for $\beta=1/4$.
As can be seen, the analytically determined locus and numerical zeros are already indistinguishable when $N=1000$. In this case, the critical curve impacts the real axis at an angle $\pi/2$, which is characteristic of a first-order transition.
The density of zeros at this first-order phase transition point is recovered using Equation \eqref{mus}. If we parameterize the curve $\gamma$ by the distance from the critical point $|\Delta \alpha| = |\alpha - \alpha_{cr}|$, which
is located at $\alpha_{cr}=\beta=1/4$, the density at $\alpha_{cr}=1/4$ 
is finite $\mu_{cr} = 4/(3\pi) > 0$.


\section{Discussion}

As well as being  calculated numerically from $Z_N(\alpha,\beta)$ in \cite{bePRL} the analytical locus of zeros for the ASEP normalization has also been obtained previously by analytic
calculations of ``free energy'' for the ASEP \cite{bwjk2} 
\begin{equation}
F =  \lim_{N \to \infty} \frac{1}{N} Z_N (\alpha,\beta).
\end{equation}
{It} 
 is interesting to compare the locus of zeros for the ASEP normalization obtained in \cite{bwjk2} with the results here 
(e.g., \eqref{psid}). In \cite{bwjk2} it was found that
\begin{small}
\begin{equation}
\label{eq:fsdef}
\begin{array}{llll}
F_1 = & - \ln [\alpha ( 1 - \alpha)]\;, & \textrm{for} \
\beta>\alpha, \; \alpha<1/2 \; , & \textrm{region} \, (1) \\
F_2 = & - \ln [\beta ( 1 - \beta )]\;, & \textrm{for} \ 
\alpha > \beta , \; \beta<1/2 \; , & \textrm{region} \, (2) \\
F_3 = & - \ln [ 1/4] \;, & \textrm{for} \ 
\alpha>1/2, \; \beta> 1/2\;, & \textrm{region} \, (3)\\
{}  
\end{array}
\end{equation}
\end{small}

\noindent
where the corresponding regions in Figure \ref{ASEPphase} are indicated. 
Assuming that this ``free energy'' derived from ASEP normalization
can be treated like the free energy of an equilibrium model, which can be justified post hoc by the Dyck path equivalence, allowing a direct determination of the locus of partition function zeros since these mark the boundaries of the different phases.

Using the electrostatic analogy described in the Introduction, the loci of the zeros are determined by observing that $\Re \psi_i =
\Re \psi_j$ for adjacent phases $i,j$. The loci of the zeros are also determined by matching the real parts of the different analytic expressions for the complex free energy in different phases in \cite{bbckk1, bbckk2}. 
In the case of the ASEP, this is equivalent to 
\begin{equation}
| \alpha ( 1 - \alpha ) | = 1/4 \; ,
\end{equation}
for the second-order transition line where $F_1=F_3$ or 
\begin{equation}
| \beta ( 1 - \beta ) | = 1/4 \; ,
\end{equation}
for the second-order transition line where $F_2=F_3$.
Across the first-order line when $\alpha=\beta<1/2$ we have $F_1=F_2$, so
\begin{equation}
| \alpha ( 1 - \alpha ) | = | \beta ( 1 - \beta ) | \; .
\end{equation}
{The} 
 locus of zeros in Figure \ref{fig:beta34} 
 is reproduced as the $f_D$ image of the circle $|z|=1/4$ 
 and it is representative of the second-order phase transition 
 line between regions (1) and (3) in Figure \ref{ASEPphase}
 (or regions (2) and (3) if the roles of $\alpha$ and $\beta$ were changed).
 The locus of zeros in Figure \ref{fig:beta14} 
 is reproduced as the $f_D$ image of the circle $|z|=\alpha(1-\alpha)=\beta(1-\beta)<1/4$ and is representative of the first-order phase transition line between regions (1) and (2).
 
 This ``physical''/electrostatic, rather than analytical, approach to obtaining the loci of zeros also recovers the results for the random allocation model outlined in Section \ref{sec:GC}. There the electrostatic potential $\Phi(u) = \Re \psi(u)$ plays the role of the thermodynamic potential $F$, so matching the real parts of this for the fluid and condensed phases in Equation \eqref{psi_solution} gives $|z_0(u)| =1$. Inverting this using Equation \eqref{pole} gives the locus of zeros in the $u$ plane found in \cite{us}, i.e., $u = 1/f(\sigma e^{i s} )$.

To summarize, we have obtained (analytically) the locus of the Yang--Lee zeros for various lattice path models and the ASEP normalization or, equivalently, for the partition function of an adsorbing one-transit Dyck walk. For the random allocation models, the path models and the ASEP the loci of zeros and their density are straightforwardly given by conformal mappings of a circle. The required mappings are determined from a generating function $f(z) = {\sum}_{s=1}^\infty w(s) z^s$ for random allocation models and 
$f_D(z)=(1 - \sqrt{1 - 4 z }) /2$ for adsorbing Dyck walks and ASEP. This follows a contour integral evaluation of the normalization/partition functions using the grand-canonical generating functions as integrands.  The results obtained for the ASEP and the random allocation model are consistent with those obtained by matching the real parts of free energies (the electrostatic analogy) and with the numerically calculated zeros.

It would be interesting to explore whether obtaining the exact locus of zeros from the conformal map technique used here could be extended to models with {\it pair}-factorized partition functions/steady-state normalizations or urn-type models with multiple constraints.

\acknowledgments{
We would like to thank M. Kieburg for discussions and contributing some of the key ideas.
In addition, it should be noted that Ralph Kenna contributed greatly to DAJ’s understanding of partition function zeros and finite size
scaling, through joint works such as \cite{wjk} and elsewhere. He will be sadly missed.}


\end{document}